\documentclass[fleqn,usenatbib,useAMS]{mnras}
\usepackage{graphicx}	
\usepackage{amssymb}
\usepackage{amsmath}
\usepackage{multicol}
\usepackage{bm}
\usepackage{pdflscape}
\usepackage{hyperref}
\usepackage{cleveref}

\DeclareMathOperator*{\argmax}{argmax}

\crefformat{figure}{Fig.~#2#1#3}
\crefformat{equation}{equation~(#2#1#3)}
\crefformat{section}{section~#2#1#3}
\crefformat{table}{Tab.~#2#1#3}
\crefformat{appendix}{appendix~#2#1#3}
\crefmultiformat{figure}{Figs.~#2#1#3}{~and~#2#1#3}%
    {,~#2#1#3}{,~#2#1#3}
\crefrangeformat{figure}{Figs.~(#3#1#4--#5#2#6)}

\newcommand{\comment}[1]{}

\usepackage[T1]{fontenc}
\usepackage{ae,aecompl}
\usepackage{newtxtext,newtxmath}

\usepackage[colorinlistoftodos]{todonotes}

\title[\textsc{margarine}: marginal Bayesian statistics]{Marginal Post Processing of Bayesian Inference Products with Normalizing Flows and Kernel Density Estimators}

\author[H. T. J. Bevins et al.]{Harry T. J. Bevins$^{1, 2}$,
\thanks{htjb2@cam.ac.uk}
William J. Handley$^{1, 2}$,
Pablo Lemos$^{3, 4}$,
Peter H. Sims $^{5}$,
Eloy de Lera Acedo $^{1, 2}$,
\newauthor
Anastasia Fialkov $^{2, 6}$,
Justin Alsing$^{7}$ \\
$^1$ Astrophysics Group, Cavendish Laboratory, Cambridge, CB3 0HE, UK \\
$^{2}$ Kavli Institute for Cosmology, Cambridge, CB3 0HA, UK \\
$^{3}$ Department of Physics \& Astronomy, University College London, London, WC1E 6BT, UK \\
$^{4}$ Department of Physics and Astronomy, University of Sussex, Brighton, BN1 9QH, UK \\
$^{5}$ Department of Physics and Trottier Space Institute, McGill University, \\ Montreal, QC H3A 2T8, Canada \\
$^{6}$ Institute of Astronomy, University of Cambridge, Cambridge CB3 0HA, UK \\
$^{7}$ Oskar Klein Centre for Cosmoparticle Physics, Department of Physics,\\ Stockholm University, Stockholm SE-106 91, Sweden}

\date{Last updated 2020 June 10; in original form 2013 September 5}

\pubyear{2020}

\begin{document}
\label{firstpage}
\pagerange{\pageref{firstpage}--\pageref{lastpage}}
\maketitle

\begin{abstract}
Bayesian analysis has become an indispensable tool across many different cosmological fields including the study of gravitational waves, the Cosmic Microwave Background and the 21-cm signal from the Cosmic Dawn among other phenomena. The method provides a way to fit complex models to data describing key cosmological and astrophysical signals and a whole host of contaminating signals and instrumental effects modelled with `nuisance parameters'. In this paper, we summarise a method that uses Masked Autoregressive Flows and Kernel Density Estimators to learn marginal posterior densities corresponding to core science parameters. We find that the marginal or `nuisance-free' posteriors and the associated likelihoods have an abundance of applications including; the calculation of previously intractable marginal Kullback-Leibler divergences and marginal Bayesian Model Dimensionalities, likelihood emulation and prior emulation. We demonstrate each application using toy examples, examples from the field of 21-cm cosmology and samples from the Dark Energy Survey. We discuss how marginal summary statistics like the Kullback-Leibler divergences and Bayesian Model Dimensionalities can be used to examine the constraining power of different experiments and how we can perform efficient joint analysis by taking advantage of marginal prior and likelihood emulators. We package our multipurpose code up in the pip-installable code \textsc{margarine} for use in the wider scientific community.
\end{abstract}

\begin{keywords}
methods: statistics, methods: data analysis, cosmic background radiation, dark ages, reionization, first stars
\end{keywords}

%%%%%%%%%%%%%%%%%%%%%%%%%%%%%%%%%%%%%%%%%%%%%%%%%%

%%%%%%%%%%%%%%%%% BODY OF PAPER %%%%%%%%%%%%%%%%%%

\section{Introduction}
\label{sec:intro}

Bayesian analysis is a cornerstone of modern statistical inference. It uses Bayes theorem to iteratively update the probability of a given hypothesis or model (posterior). The inference is informed with existing knowledge about the model parameters (prior) and we define a representative probability distribution for the data given the choice of model and parameters (likelihood) to perform the inference. Various computational approaches to Bayesian analysis exist, including Markov Chain Monte Carlo~(MCMC) methods \citep[e.g.][]{emcee} and Nested Sampling algorithms \citep[e.g.][]{polychord_2015, polychord_2015b}.

The technique has been applied in the inference of cosmological parameters from experiments such as the CMB mapper Planck~\citep{Planck_cosmo_2020}, the Dark Energy Survey~\citep[DES,][]{DES_Year1_2018, DES_year3_2021, DES_year3}, 21-cm power spectrum experiments such as HERA~\citep{HERA}, LOFAR~\citep{Ghara_LOFAR_2020, Mondal_LOFAR_2020} and MWA~\citep{Greig_MWA_2020, Ghara_MWA_2021} and global (sky-averaged) 21-cm experiments, to infer constraints on the statistical properties of early galaxy populations, such as REACH~\citep{Anstey_REACH_2021, Acedo_reach_mission_2022} and SARAS \citep{Bevins_SARAS2_2022, Bevins_SARAS3_2022}. 

These experiments are often plagued by `nuisance' parameters that characterise foregrounds and instrumental effects in the data. This leads to high dimensional parameter spaces, of which only a few of the parameters model the core science. Consequently, drawing conclusions about the parameters of interest and making comparisons across different experiments with different nuisance parameters can be challenging. For example, in the Year 1 analysis of data from DES the likelihood contains a total of 26 parameters of which 20 could be considered nuisance parameters with only 6 corresponding to cosmological parameters of interest~\citep{DES_Year1_2018}. Similarly, for REACH the likelihood can contain upwards of 15 parameters, with only 3-7 of these corresponding to the cosmological 21-cm signal.

We present a post-processing tool called \textsc{margarine} that can be used to calculate \emph{marginal}, \emph{nuisance-free} or equivalently \emph{nuisance-marginalised} posterior distributions pertaining to the core science goals of the above experiments through density estimation with Masked Autoregressive Flows~\citep[MAFs,][]{Papamarkarios_MAF_2017} and Kernel Density Estimators~\citep[KDEs,][]{rosenblatt_KDE_1956, parzen_KDE_1962}.

Density estimators model probability distributions given a set of representative samples, and there are a number of different methods through which this can be done.They can be used to evaluate the logarithm of the probability of a set of samples on the learned distribution, meaning that they can be used as a computationally inexpensive likelihood generators and, when trained on marginal samples from sub-spaces of a larger distribution, to calculate marginal statistics such as the marginal Kullback-Leibler divergence and marginal Bayesian Model Dimensionality. The density estimators used in this work are outlined in more detail in \cref{sec:density_estimators}.

As likelihood generators, the density estimators can be used to perform efficient joint analysis of constraints from multiple different experiments probing the same core science with different nuisance parameters.

The Kullback-Leibler~(KL) divergence can be used to determine how much information has been gained about the parameters of a model through our Bayesian analysis when moving from the prior and the posterior \citep{kullback_information_1951}. The marginal Kullback-Leibler~(KL) divergence, accessible through the density estimators discussed above, is a measure of how much information is gained in the inference when contracting specifically the core science prior onto the core science posterior without including contributions from correlations with or constraints on the nuisance parameters. This allows us to confidently compare the information gained from observations by different experiments of the same signal, regardless of their specific instrumental nuances. 

The Bayesian Model Dimensionality gives a measure of the number of parameters that have been constrained during our inference \citep{Handley_dimensionality_2019}. The marginal Bayesian Model Dimensionality~(BMD) can be used to compare the effective number of core science parameters constrained by different experiments and models independent of the experiments specific nuisance parameters. The KL divergence and BMD are further explored in \cref{sec:theory}.

Both the \emph{marginal} KL divergence and the BMD, which prior to \textsc{margarine} were not easy to assess, can therefore provide a way to quantitatively determine which experimental approaches are the most informative, leading to improvements to instrumentation in the future.

Finally, we note that if the density estimator is set up correctly such that it is a bijective transformation between the unit hyper-cube and the target posterior, then it can be used to generate the prior on a subsequent Nested Sampling or MCMC run. This allows far more complex priors based on current experimental results to be used, incorporating our current knowledge of the parameter space~\citep[this idea was first introduced in ][]{Alsing_bijectors_2021}.

In addition to the previously discussed examples, Bayesian analysis has historically been applied to many different sub-fields in astrophysics, such as the study of velocity fields \citep[e.g][]{Kaiser_Lahav_1989}, the CMB \citep[e.g.][]{Trotta_bayes_2008}, the study of galaxies from JWST \citep[e.g.][]{curtis-lake_jwst_2023} and gravitational wave studies \citep[e.g.][]{Romero-shaw_gw_2021, nanograv_2023} among others. Due to the ubiquity of Bayesian analysis in cosmology and astrophysics, post-processing of samples from MCMC and Nested Sampling algorithms is an important area of research. \textsc{margarine}, presented in this paper, is an invaluable addition to the Bayesian workflow and allows for direct and specific comparison of the constraining ability of different experimental approaches, rapid emulation of nuisance-free likelihoods and experimentally informed priors.

We implement our approach in \textsc{Python} using \textsc{tensorflow} and the \textsc{keras} backend and release the code as the pip installable package \textsc{margarine}. The code is fully documented with examples, subject to continuous integration testing and generally does not require high performance computing to be used.

In \cref{sec:theory} we introduce Bayesian analysis and the marginal statistical quantities that can be evaluated with \textsc{margarine}. We discuss the two different types of density estimators used by \textsc{margarine} in \cref{sec:density_estimators}. In \cref{sec:applications}, we demonstrate several applications of the code base, including applications to samples from the Dark Energy Survey and mock observations with the REACH pipeline. We summarise the paper in \cref{sec:conclusions}.

\textsc{margarine} was previously introduced in the conference proceedings for the 2022 International Conference on Bayesian and Maximum Entropy Methods in Science and Engineering \citep[MaxEnt22, ][]{margarine_maxent} in which the nuisance-free likelihood was derived and demonstrated. We review this work briefly in this paper.

\section{Theory}
\label{sec:theory}

\subsection{Bayesian analysis}
\label{sec:bayesian_inference}

Bayesian analysis is concerned with the estimation of posterior probabilities using Bayes theorem
\begin{equation}
    P(\Theta | D, M) = \frac{P(D| \Theta, M) P(\Theta|M)}{P(D|M)} = \frac{\mathcal{L}(\Theta)\mathcal{\pi}(\Theta)}{\mathcal{Z}},
    \label{eq:bayes_theorem}
\end{equation}
where $\Theta$ are the parameters of our model $M$ describing the data $D$ and $\mathcal{L}(\Theta)$ corresponds to the likelihood. The likelihood, the probability of the data given the model, is often assumed to be Gaussian in nature \citep[although more complicated likelihoods can be implemented,][]{Scheutwinkel2022a}. 
The prior, $\mathcal{\pi}(\Theta) = P(\Theta|M)$, encodes our existing knowledge about the parameters before any sampling is performed. $\pi(\Theta)$ is often chosen to be uniform or log-uniform for each parameter within some physically motivated range, however it can be useful to inform this with existing experimental results.

The evidence, $\mathcal{Z}$, is a marginal likelihood integrated over all the parameters weighted by the prior
\begin{equation}
    \mathcal{Z} = \int \mathcal{L}(\Theta)\mathcal{\pi}(\Theta)d\Theta.
    \label{eq:evidence}
\end{equation}
The estimation of $\mathcal{Z}$ via \cref{eq:evidence} is the primary concern of the Nested Sampling algorithm~\citep{skilling_nested_2004} and through the evaluation of \cref{eq:evidence} we can derive samples on the posterior $\mathcal{P}(\Theta| D, M)$. The algorithm works by generating samples in $\Theta$ from the prior, calculating the likelihood and numerically approximating the integral in \cref{eq:evidence}. At each iteration, new samples at higher likelihoods are generated to sample the posterior bulk. The method by which the samples are generated is different in different implementations~\citep[e.g.][]{multinest_2009, polychord_2015}, and the efficiency of the sampling algorithm is often dependent on the dimensionality of the parameter space. The evidence can be used for model comparison, with a higher evidence indicating a preference for one model over another.

If our model $M (\Theta)$ contains nuisance, $\alpha$, and cosmological signal parameters, $\theta$, then we can write $\Theta = \{\alpha, \theta\}$. Mathematically, the marginal posterior for the signal parameters, $\mathcal{P}(\theta)$, can be calculated by integrating out the dependence on $\alpha$
\begin{equation}
    \mathcal{P}(\theta| D, M) = \int \mathcal{P}(\alpha, \theta|D, M) d \alpha.
\end{equation}

Whilst we have access to samples in the marginal space, however, the probability $\mathcal{P}(\theta|D, M)$, referred to as $\mathcal{P}(\theta)$ from now on, is typically a difficult quantity to calculate as it involves performing a high dimensional integral over the nuisance parameters $\alpha$. We can effectively perform this marginalisation by training density estimators on samples in $\theta$ from MCMC or Nested Sampling chains excluding the samples in $\alpha$. From these density estimators, we can generate samples from the distribution $\mathcal{P}(\theta)$ and importantly calculate $\log \mathcal{P}(\theta)$ for a given set of $\theta$ which gives us access to a whole host of previously intractable marginal summary statistics. The density estimators used in this work are outlined in more detail in \cref{sec:density_estimators}.

\subsection{Nuisance-free likelihood}
\label{sec:nuisance-free-likelihood}

In \cite{margarine_maxent} we defined the `nuisance-free' likelihood function as
\begin{equation}
    \mathcal{L}(\theta) 
\equiv \frac{\int\mathcal{L}(\theta,\alpha)\pi(\theta,\alpha)d\alpha}{\int \pi(\theta,\alpha)d\alpha} = \frac{\mathcal{P}(\theta)\mathcal{Z}}{\pi(\theta)},
    \label{eqn:partial}
\end{equation}
where the marginal posterior, $\mathcal{P}(\theta)$, and prior, $\pi(\theta)$, can be emulated from samples with a density estimator in \textsc{margarine}. Here $\mathcal{Z}$ is the Bayesian evidence from the full fit with $\theta$ and $\alpha$ and is crucial to the definition of the `nuisance-free' likelihood. Without the inclusion of the evidence, any resultant sampling performed with $\mathcal{L}(\theta)$ will not produce sensible evidences that can be used for model comparison.

We demonstrate applications of the nuisance-free likelihood to perform efficient joint analysis in \cref{sec:nuisance_free_likelihood_apps} with a simple toy example and reference \cite{margarine_maxent} for a more complete description.

\subsection{Marginal Kullback-Leibler divergence}

As mentioned in the introduction, the KL divergence quantifies the information gain when moving from the prior to posterior and is an effective measure of the contraction between the two distributions. The \emph{marginal} KL divergence corresponding to the astrophysical or cosmological parameters $\theta$ is defined as the average Shannon Information
\begin{equation}
    \mathcal{I}(\theta) = \log\bigg(\frac{\mathcal{P}(\theta)}{\mathcal{\pi}(\theta)}\bigg)
    \label{eq:shannon_entropy}
\end{equation}
over the posterior, $\mathcal{P}(\theta)$,
\begin{equation}
    \mathcal{D}(\mathcal{P}||\pi) = \int \mathcal{P}(\theta) \log\bigg(\frac{\mathcal{P}(\theta)}{\mathcal{\pi}(\theta)}\bigg) d\theta = \left\langle \mathcal{I} \right\rangle_\mathcal{P} \approx \log \bigg(\frac{V_\pi}{V_\mathcal{P}}\bigg),
    \label{eq:kl_divergence}
\end{equation}
and is a measure of how much information in bits the data provides us about the astrophysical or cosmological part of the parameter space. It is approximately equal to the log of the ratio of the prior volume,$V_\pi$, and the posterior volume, $V_\mathcal{P}$. $\mathcal{D}$ is a strong function of the prior, and inherits the property of being additive for independent parameters from the Shannon Information. Calculation of the KL divergence requires the posterior distribution to be appropriately normalized, which requires knowledge of the Bayesian evidence, and consequently the quantity is not attainable with common MCMC sampling techniques which generate samples on the non-normalised posterior through the relationship
\begin{equation}
    \mathcal{P}(\Theta | D, M) \propto \mathcal{L}(\Theta) \pi(\Theta).
\end{equation}

The KL-divergence can be related to the Bayesian evidence via
\begin{equation}
	\log \mathcal{Z} = \langle \log \mathcal{L} \rangle - \mathcal{D},
\end{equation}
and acts as an Occam penalty that penalizes complex models with large numbers of dimensions \citep{Hergt2021}. In practice, the above equation is typically used to calculate the KL divergence from Nested Samples, where we have ready access to the Bayesian evidence $\mathcal{Z}$. However, using density estimators to calculate the KL divergence, marginal or not, also has the added advantage that it does not require knowledge of the Bayesian evidence because the density estimators naturally return normalized distributions.

\subsection{Marginal Bayesian Model Dimensionality}

An alternative measure of constraint is the Bayesian Model Dimensionality~\citep[BMD,][]{Handley_dimensionality_2019} which gives a measure of the effective number of parameters that are being constrained and can be used to quantify tensions between different experimental results~\citep{Handley_tensions_2019,2022arXiv220505892G}. The BMD is defined such that a Gaussian constraint around a single parameter corresponds to a value of one, as does a correlated constraint between two parameters. This is demonstrated visually in \cite{Handley_dimensionality_2019} and allows us to predict the expected BMD for a given posterior by asking how Gaussian the distribution is.

The \emph{marginal} BMD $d$ is given by
\begin{equation}
    \begin{split}
    \frac{d}{2} &= \int \mathcal{P}(\theta) \bigg(\log\bigg(\frac{\mathcal{P}(\theta)}{\mathcal{\pi}(\theta)}\bigg) - \mathcal{D}\bigg)^2 d\theta \\
    & = \mathrm{var}(\mathcal{I})_\mathcal{P} \\
    & = \langle(\log \mathcal{L})^2\rangle_\mathcal{P} - \langle\log\mathcal{L}\rangle_\mathcal{P}^2,
    \end{split}
    \label{eq:bayesian_dimensionality}
\end{equation}
and a full derivation of the BMD can be found in~\cite{Handley_dimensionality_2019}. 
The quantity is the variance of the Shannon information and therefore a higher order statistic than the KL divergence.
It is only weakly prior dependent and like the KL divergence is additive for independent parameters and invariant under a change of variables.

\section{Density Estimators in Practice}
\label{sec:density_estimators}

In principle, one can use any type of density estimator to model subspaces in a larger parameter space and calculate the marginal Bayesian statistics discussed. The only requirements are that it can be used to resample the space and approximate the \textit{normalised} log-probability associated with the posterior subspace.

\subsection{Masked Autoregressive Flows}

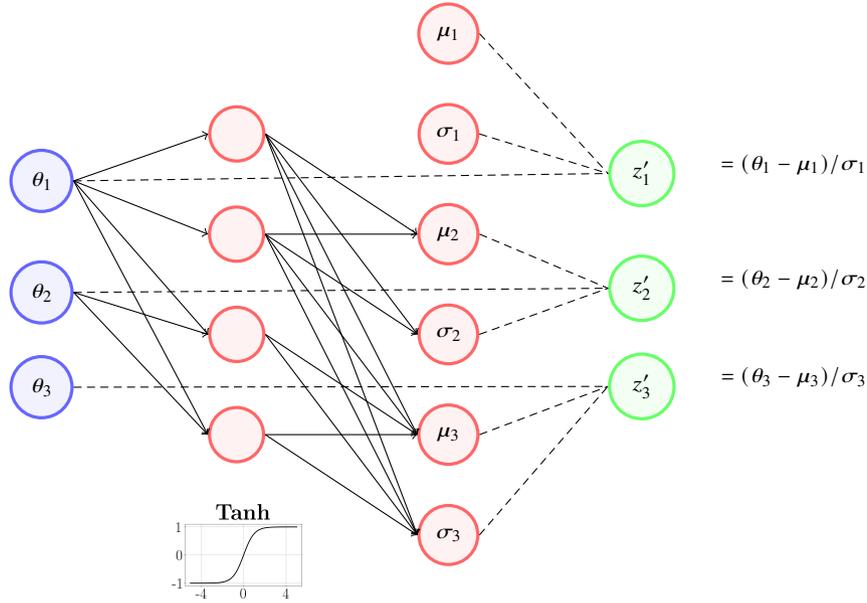
\begin{figure*}
    \centering
    \begin{tikzpicture}[rednode/.style={circle, draw=red!60, fill=red!5, very thick, minimum size=5mm},
                bluenode/.style={circle, draw=blue!60, fill=blue!5, very thick, minimum size=5mm},
                greennode/.style={circle, draw=green!60, fill=green!5, very thick, minimum size=5mm},
                node distance=0.5cm and 2cm,
                remember picture]
        \node (0) {};
        
        \node[rednode, above=of 0, text width=0.5cm, align=center](layer1_center1) {$\mu_2$};
        \node[rednode, below=of layer1_center1, text width=0.5cm, align=center](layer1_center2) {$\sigma_2$};
        \node[rednode, above=of layer1_center1, text width=0.5cm, align=center](layer1_top2) {$\sigma_1$};
        \node[rednode, above=of layer1_top2, text width=0.5cm, align=center](layer1_top1) {$\mu_1$};
        \node[rednode, below=of layer1_center2, text width=0.5cm, align=center](layer1_bottom1) {$\mu_3$};
        \node[rednode, below=of layer1_bottom1, text width=0.5cm, align=center](layer1_bottom2) {$\sigma_3$};
    
        \node[rednode, left=of layer1_top2, text width=0.5cm, align=center](hl1) {};
        \node[rednode, left=of layer1_center1, text width=0.5cm, align=center](hl2) {};
        \node[rednode, left=of layer1_center2, text width=0.5cm, align=center](hl3) {};
        \node[rednode, left=of layer1_bottom1, text width=0.5cm, align=center](hl4) {};
        
        \node[below=of hl4, inner sep=0pt] (tanh) {\includegraphics[width=.1\textwidth]{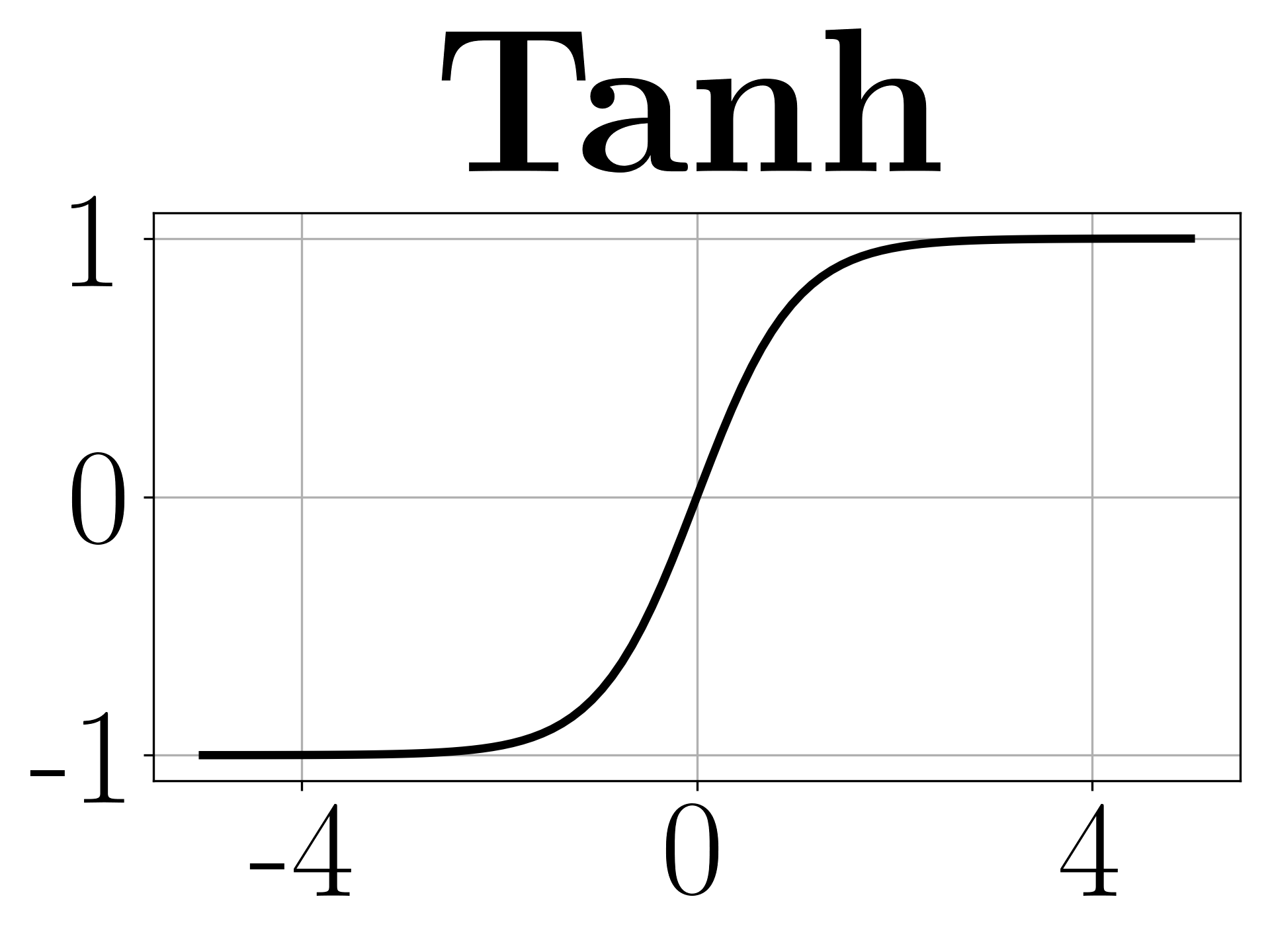}};
        
        \node[bluenode, below left=of hl2, text width=0.5cm, align=center, yshift=0.3cm](input_2) {$\theta_2$};
        \node[bluenode, above=of input_2, text width=0.5cm, align=center, yshift=0.12cm](input_1) {$\theta_1$};
        \node[bluenode, below=of input_2, text width=0.5cm, align=center, yshift=0.1cm](input_3) {$\theta_3$};

        \node[greennode, right=of 0, text width=0.5cm, align=center, yshift=0.3cm](output_2) {$z^\prime_2$};
        \node[greennode, above=of output_2, text width=0.5cm, align=center, yshift=0.12cm](output_1) {$z^\prime_1$};
        \node[greennode, below=of output_2, text width=0.5cm, align=center, yshift=0.1cm](output_3) {$z^\prime_3$};
        
        \node[right=of output_2, align=left, xshift=-1.5cm, yshift=0.12cm] {$= (\theta_2 -\mu_2) / \sigma_2$};
        
        \node[right=of output_1, align=left, xshift=-1.5cm, yshift=0.12cm] {$= (\theta_1 -\mu_1) / \sigma_1$};
        
        \node[right=of output_3, align=left, xshift=-1.5cm, yshift=0.12cm] {$= (\theta_3 -\mu_3) / \sigma_3$};
        
        \draw[densely dashed](input_1.east) -- (output_1.west);
        
        \draw[densely dashed](input_2.east) -- (output_2.west);
        
        \draw[densely dashed](input_3.east) -- (output_3.west);
        
        \draw[->](input_1.east) -- (hl1.west);
        \draw[->](input_1.east) -- (hl2.west);
        \draw[->](input_1.east) -- (hl3.west);
        \draw[->](input_1.east) -- (hl4.west);
        
        %\draw[->](input_2.east) -- (hl1.west);
        %\draw[->](input_2.east) -- (hl2.west);
        \draw[->](input_2.east) -- (hl3.west);
        \draw[->](input_2.east) -- (hl4.west);
        
        %\draw[->](input_3.east) -- (hl1.west);
        %\draw[->](input_3.east) -- (hl2.west);
        %\draw[->](input_3.east) -- (hl3.west);
        %\draw[->](input_3.east) -- (hl4.west);
        
        \draw[->](hl1.east) -- (layer1_center1.west);
        \draw[->](hl2.east) -- (layer1_center1.west);
        %\draw[->](hl3.east) -- (layer1_center1.west);
        %\draw[->](hl4.east) -- (layer1_center1.west);
        
        \draw[->](hl1.east) -- (layer1_center2.west);
        \draw[->](hl2.east) -- (layer1_center2.west);
        %\draw[->](hl3.east) -- (layer1_center2.west);
        %\draw[->](hl4.east) -- (layer1_center2.west);
        
        %\draw[->](hl1.east) -- (layer1_top1.west);
        %\draw[->](hl2.east) -- (layer1_top1.west);
        %\draw[->](hl3.east) -- (layer1_top1.west);
        %\draw[->](hl4.east) -- (layer1_top1.west);
        
        %\draw[->](hl1.east) -- (layer1_top2.west);
        %\draw[->](hl2.east) -- (layer1_top2.west);
        %\draw[->](hl3.east) -- (layer1_top2.west);
        %\draw[->](hl4.east) -- (layer1_top2.west);
        
        \draw[->](hl1.east) -- (layer1_bottom2.west);
        \draw[->](hl2.east) -- (layer1_bottom2.west);
        \draw[->](hl3.east) -- (layer1_bottom2.west);
        \draw[->](hl4.east) -- (layer1_bottom2.west);
        
        \draw[->](hl1.east) -- (layer1_bottom1.west);
        \draw[->](hl2.east) -- (layer1_bottom1.west);
        \draw[->](hl3.east) -- (layer1_bottom1.west);
        \draw[->](hl4.east) -- (layer1_bottom1.west);
        
        \draw[densely dashed](layer1_top1.east) -- (output_1.west);
        \draw[densely dashed](layer1_top2.east) -- (output_1.west);
        
        \draw[densely dashed](layer1_center1.east) -- (output_2.west);
        \draw[densely dashed](layer1_center2.east) -- (output_2.west);
        
        \draw[densely dashed](layer1_bottom1.east) -- (output_3.west);
        \draw[densely dashed](layer1_bottom2.east) -- (output_3.west);

    \end{tikzpicture}
    \caption{The Masked Autoencoder for Distribution Estimation~(MADE) architecture, when trained appropriately, transforms samples from a complex probability distribution $P(\theta)$ on to samples from a standard normal distribution under the assumption that the probability distribution $P(\theta)$ can be broken into conditional one-dimensional Gaussian probability distributions. Many different MADE networks can be stacked together and trained to produce a normalising flow increasing the expressivity of the density estimation. By definition, normalising flows are bijective, and a trained implementation can be used to calculate $\log P(\theta)$ and draw samples from $P(\theta)$. At each hidden layer in the MADE we use a `tanh' activation function.}
    \label{fig:maf}
\end{figure*}

Complex densities are best estimated using expressive neural networks \citep{Papamakarios2019} that have been trained to transform between some base distribution and the target probability distribution. \textsc{margarine} uses Masked Autoregressive Flows \citep[MAF,][]{Papamarkarios_MAF_2017} to perform this transformation.

An example Masked Autoencoder for Distribution Estimation~(MADE) neural network, which forms the basis of the MAF, is shown in \cref{fig:maf}. The network works by dividing the target probability distribution into a series of one dimensional conditional distributions
\begin{equation}
P(\theta) = \prod_i P(\theta_i| \theta_1, \theta_2, ..., \theta_{i-1}),
\label{eq:conditional_1}
\end{equation}
and approximating each conditional probability as a Gaussian
\begin{equation}
P(\theta_i| \theta_1, \theta_2, ..., \theta_{i-1}) = \mathcal{N}(\mu_i, \sigma_i)
\label{eq:conditional_2}
\end{equation}
where $\sigma_i$ and $\mu_i$ are the standard deviation and mean of the conditionals. Throughout the paper, we use subscripts for dimensions and superscripts for samples unless otherwise stated. The ability to effectively describe the target distribution with \cref{eq:conditional_2} is a standard assumption \citep{Papamarkarios_MAF_2017, Papamakarios2019} with known failure modes that have been extensively reviewed in the literature \citep{izmailov2020semi, ardizzone2020training, hagemann2021stabilizing, Stimper2021}. A discussion of the failure modes is beyond the scope of this paper but explored in the context of \textsc{margarine} in \citep{bevins_pnf_2023}. Typically, every node in a neural network between the input and output are connected in what we would call a fully connected neural network. In a Masked Autoregressive Flow we mask some of these connections to represent the conditionality in \cref{eq:conditional_1} as is shown in \cref{fig:maf}. Due to the Gaussian nature of our model for the target distribution, we typically use a multivariate standard normal distribution, $z \sim \mathcal{N}(0, 1)$, as our base distribution.

We input samples, $j$, from $\theta$ into the network and output values of $\sigma$ and $\mu$ which are then used to reconstruct samples on the standard normal distribution, $z$, in dimension $i$ with
\begin{equation}
    z_i^{j,\prime} = \frac{(\theta_i^j - \mu_i^j)}{\sigma_i^j}.
    \label{eq:reconstruction_maf}
\end{equation}
The vectors of $\sigma$ and $\mu$ are therefore functions of the networks weights and biases that need to be trained.

We train the weights and biases using the change of variables formula
\begin{equation}
    P(x) = P(z) \bigg|\frac{\delta z}{\delta \theta}\bigg|,
\end{equation}
where the goal is to minimize the KL-divergence between the true probability distribution of the samples $\theta$ and the prediction from the network $P_w(\theta)$ given by
\begin{equation}
    \mathcal{D}(P(\theta)||P_{w}(\theta)) = -\mathbb{E}_{P(\theta)}[\log P_{w}(\theta)] + \mathbb{E}_{P(\theta)}[\log P(\theta)],
\end{equation}
where $w$ are the weights of the network and $\mathbb{E}_{P(\theta)}$ is the expectation value over $P(\theta)$. The second term in the KL divergence is independent of the network, and so it is constant when we change the hyperparameters of the network. We can therefore focus on minimizing the first term of the KL divergence
\begin{equation}
    -\mathbb{E}_{P(\theta)}[\log P_{w}(\theta)] = -\frac{1}{N} \sum_{i=0}^N \log P_{w}(\theta_i),
\end{equation}
meaning that our minimization problem becomes a maximization problem over the log-probability predicted by the network for the samples on the target distribution 
\begin{equation}
    \argmax_{w} \sum^N_{j=0} \log P_{w}(\theta^j).
\end{equation}
Equivalently by a change of variables
\begin{equation}
    \argmax_{w} \sum_{j=0}^N [\log P_{w}(z^{\prime j}) + \log \bigg|\bigg(\frac{\delta z^{\prime}}{\delta \theta}\bigg|_{z^{\prime j}, \theta^j}\bigg)\bigg| ],
\end{equation}
where $z^\prime$ is a function of the network weights $w$ \citep{Alsing2019}. Here the second term is just the derivative over the network and the first term is trivially calculated from the output $\sigma$ and $\mu$.

Importantly, the MADE networks are bijective, meaning that there is a one-to-one relationship between samples in the base distribution and target distribution. We can therefore, once trained, use them to transform samples back and forth between the two.

By chaining a series of these networks together, we create a MAF. This allows the emulator to learn more complex distributions and can be expressed using the following formalism
\begin{equation}
    \begin{split}
    z_0 & = \mathcal{N}(0, 1) \\
    z_1 & = z_0 \sigma_1(z_0, w_1) + \mu_1(z_0, w_1)\\
    \vdots &\\
    \theta & = z_{n-1} \sigma_{n}(z_{n-1}, w_{n-1}) + \mu_{n}(z_{n-1}, w_{n-1})
    \end{split}
\label{eq:MAF}
\end{equation}
where the index here refers to a given network in the chain. We train the chain of neural networks simultaneously with one loss function and feed the outputs of proceeding networks into subsequent networks in the chain. 

\begin{figure*}
    \centering
    \includegraphics{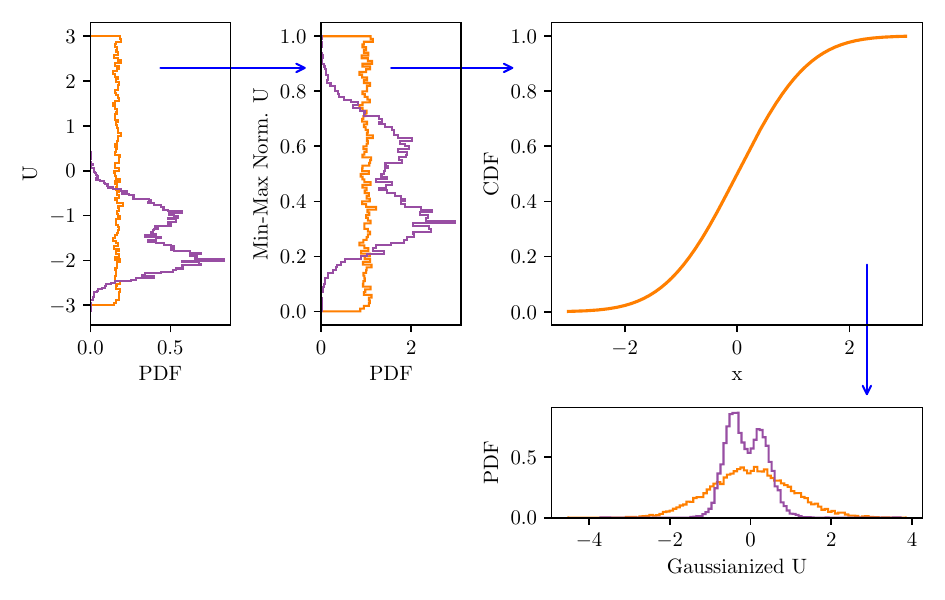}
    \caption{The figure demonstrates how a 1D distribution is Gaussianized to aid with training of the Masked Autoregressive Flows and Kernel Density Estimators. We show the process being applied to two different distributions; a uniform distribution in orange and Gaussian-like distribution in purple. The samples are first min-max normalized so that the range of the distributions is 0-1. They are then pushed through the standard normal CDF to Gaussianize the samples. As can be seen for a uniform distribution, the resultant distribution is the standard normal. However, for the non-trivial distribution in purple, the result is a shifted and scaled version of the standard normal. %This is exactly the transformation that the Masked Autoregressive Flow is attempting to model, and is easier to learn with a kernel density estimate than the complex target in the non-normalized space. 
    This process is done in each dimension before training and samples drawn from the flows or KDEs are in the Gaussianized space but can easily be transformed back to the non-normalized space.}
    \label{fig:Gaussianized}
\end{figure*}

To improve the accuracy of the density estimation, we can transform the target posterior samples, that we input to our MAF, into a Gaussianized space via the standard normal cumulative distribution function~(CDF). The target distribution in this normalised space is a skewed and scaled version of the base distribution $z_0$ which makes the transformation easier to learn. This is demonstrated in \cref{fig:Gaussianized}.

Due to the invariance of the KL divergence and, similarly, the BMD under a change of variables, we can calculate their values in any transformed version of the original parameter space provided the prior and posterior undergo the same change of variables.

%When estimating the KL-divergence and BMD, we can train a MAF to estimate the prior log-probability. However, in the case where the prior is uniform, or can be easily manipulated along with the posterior samples into a uniform parameter space, then the prior in the Gaussian space is the standard normal distribution, $\mathcal{N}(0, 1)$.

There are a number of tunable parameters involved when designing MAFs including the number of networks, the number of layers in each network, the number of epochs and the learning rate. All of these parameters can be explored using \textsc{margarine}, and we use recommendations from a previous work~\citep{Alsing_bijectors_2021} and set these as the defaults in \textsc{margarine}.

\subsection{Kernel Density Estimation}

A one dimensional Kernel Density Estimator~(KDE) is built by summing a series of Gaussian kernels centred around each sample $\theta_j$ in the set of samples $\theta$. The KDE provides an estimate of the probability at each sample $\theta$ and where there are more samples in $\theta$ the sum will be larger representing a higher probability.
Mathematically a 1D KDE is defined as
\begin{equation}
    \mathcal{P}(\theta) = \frac{1}{nh}\sum_{j=1}^n K\bigg(\frac{\theta-\theta^j}{h}\bigg)
    \label{eq:gauss_kde}
\end{equation}
where $h$ is known as the bandwidth and is a smoothing parameter on the Gaussian kernel $K$ with a standard deviation given by the standard deviation of the $n$ samples, and in the set $\theta$. This scales to higher dimensions where $\theta$ and $\theta_j$ become $N$ dimensional vectors and $h$ becomes a 2D matrix of bandwidths. We use a multivariate Gaussian kernel where $h$ is a matrix of smoothing parameters on the covariance matrix and $\theta_j$ is equivalent to an $N$ dimensional vector of means. Since the KDE is a sum of known kernels with a known covariance matrix and set of means, $\theta_j$, then the probability distribution of the samples is trivially calculable, as is its logarithm.

When generating KDEs, we perform the same type of normalisation of our target distribution as is done with the MAFs, transforming it into a Gaussianized space via a standard normal CDF (see \cref{fig:Gaussianized}). The transformation allows the density estimator to better capture the edges of very flat and uniform posteriors.

%KDEs are not bijective transformations, but we can generate samples from them allowing us to calculate marginal KDEs.  but transforming samples from the unit hypercube to the target density, when we want to use more complex priors in our Nested Sampling runs, is not so trivial.%~(see section 3.5 in \cite{polychord_2015b}).

%To transform the unit hypercube into samples on the target distribution via the KDE we have to break the target probability distribution down into the product of conditional probabilities
%\begin{equation}
%    P(x_1, x_2, \hdots, x_n) = P(x_1) P(x_2|x_1) \hdots P(x_n | x_1, x_2, \hdots, x_{n-1}).
%\end{equation}
%We then use the inverse CDFs of each conditional probability to transform samples on the hypercube to samples on the target distribution
%\begin{equation}
%    (u_1, u_2, \hdots, u_n) \rightarrow (x_1, x_2, \hdots, x_n),
%\end{equation}
%via
%\begin{equation}
%    \begin{split}
%       x_1 & = F_1(u_1) \\
%        x_2 & = F_2(u_2;x_1) \\
%        \vdots & \\
%       x_n & = F_n(u_n;x_1, x_2, \hdots, x_{n-1})
%    \end{split}
%\end{equation}
%where $F_i$ are the conditional and marginalised inverse CDFs. 

%For a multivariate Gaussian KDE marginalisation is equivalent to ignoring components of $h$ and $x_i$ in the multivariate equivalent of \cref{eq:gauss_kde}. Then conditioning the probability distribution simply involves analytical corrections to the means and standard deviations based on the corresponding values for the $n-1$ distributions.

We can generate samples from trained KDEs and calculate marginal Bayesian statistics, however KDEs are not strictly bijective. Adapting a KDE to be bijective is a non-linear process and requires the use of a root-finding algorithm to transform from the hypercube to the KDE space. This is a process that is not currently optimized, but is fully implemented in \textsc{margarine} and the equivalent transformation using MAFs is much more computationally efficient.

There are fewer tunable parameters for the KDE, however we can change the bandwidth of the kernel. We use the Silverman~\citep{silverman2018density} approach\footnote{This is standard practice and is a built-in option in the \textsc{scipy} \href{https://docs.scipy.org/doc/scipy/reference/generated/scipy.stats.gaussian_kde.html}{Gaussian kernel} tool.} to calculate $h$ but note that this can be modified with \textsc{margarine}. In one dimension $h$ is given by
\begin{equation}
    h = \bigg(\frac{4}{3}\bigg)^{\frac{1}{5}} n^{-\frac{1}{5}}\sigma,
\end{equation}
where $\sigma$ is the standard deviation of the data and $n$ is the number of samples \cite{silverman2018density}. In the multivariate case, this just becomes
\begin{equation}
    h_{ii} = \bigg(\frac{4}{d+2}\bigg)^{\frac{2}{d+4}} n^{-\frac{2}{d+4}} \sigma_{ii},
\end{equation}
where $d$ is the number of dimensions and $h_{ij} = 0$ for $i \neq j$.

\section{Applications of \textsc{margarine}}
\label{sec:applications}

\subsection{Marginal Joint Analysis}
\label{sec:nuisance_free_likelihood_apps}

To illustrate the application of the nuisance free likelihood function, we use an example from 21-cm cosmology. The sky-averaged (global) 21-cm signal is the differential brightness temperature between the radio background, $T_r$, and the spin temperature, $T_s$, of neutral hydrogen during the Cosmic Dawn and Epoch of Reionization. $T_s$ is a statistical temperature that quantifies the ratio of the number of neutral hydrogen atoms with proton and electron spins aligned versus anti-aligned. 

The signal manifests itself as an absorption trough against the radio background when the first stars begin to form in the early universe around redshift $z\approx 20 - 40$, and Lyman-$\alpha$ coupling drives the spin temperature down to the gas kinetic temperature. Various astrophysical processes then begin to heat the gas around $z \approx 15 - 25$, particularly as the first X-ray sources form, and the coupled spin temperature is raised back up to and sometimes in excess of the radio background. UV emission then begins to ionize the neutral hydrogen around $z \approx 5 - 10$, and the signal vanishes during the Epoch of Reionization. The signal has a rest frequency of $1420.4$~MHz and is redshifted by the expansion of the universe. Since we are interested in the evolution of the signal over the range~$z \approx 5 - 50$, we use radio telescopes to observe the signal in the range $\nu \approx 30 - 250$~MHz. The global 21-cm signal is predicted to be of order $100$~mK in magnitude \citep[although there is some uncertainty in this value e.g.][]{Mesinger_2010, Barkana_2018, Reis2020, Reis2021, Munoz_2022} and is masked by foregrounds from our own Galaxy~\citep{Bernardi_galaxy_2009}, extragalactic radio sources \citep{Nitu_background_2021} and the ionosphere~\citep{Shen_ionosphere_2021}, which collectively are around five orders of magnitude brighter than the signal. To a crude approximation, the signal can be modelled with a Gaussian absorption profile. For a review of the physics of the 21-cm signal, see \cite{Furlanetto_review_2006, Mesinger_review_2019}.

We generate two different mock sky-averaged 21-cm data sets covering different bandwidths of $50- 100$ and $85 - 190$~MHz and observing different foregrounds from different locations. We call these two experiments $A$ and $B$ and model the foreground as
\begin{equation}
    T_\mathrm{fg} = a \nu^{-\beta},
\end{equation}
where $a$ is a common scale factor for both experiments and $\beta$ is set as $-2.6$ and $-2.5$. The model is based on the expectation that the foreground is dominated by synchrotron emission and the power law is based on observational constraints from experiments like EDGES~\citep{Mozdzen_EDGES_spectral_index_2017}, SARAS3~\citep{SARAS3}, and LEDA \citep{LEDA_spectral_Index_2021}.

It is typical in analysis of data from 21-cm experiments to model the foreground as an unconstrained polynomial function due to its smooth properties. However, the structure of the antenna beam can introduce non-smooth chromatic features into the data and more complex forward modelling of the foreground is need \citep{Anstey_REACH_2021}. For simplicity, we assume an achromatic beam and thus a data set that includes a smooth foreground and Gaussian 21-cm signal. The signal in our mock data is given by
\begin{equation}
    T_{21} = - T_\mathrm{min} \exp\bigg(-\frac{(\nu - \nu_c)^2}{2\Delta \nu ^2}\bigg),
    \label{eq:Gaussian_signal}
\end{equation}
where $T_\mathrm{min}$ is the signal amplitude, $\nu$ is the frequency range of the data, $\nu_c$ is the central frequency of the absorption feature and $\Delta \nu$ is the signal's width. We set $T_\mathrm{min} = 0.25$~K, $\nu_c = 80$~MHz and $\Delta \nu = 10$~MHz.

We model each data set separately using the Nested Sampling algorithm \textsc{polychord}, \cref{eq:Gaussian_signal} and a log-log polynomial foreground model given by
\begin{equation}
    \log_{10}(T_\mathrm{fg}) = \sum^{N}_{k=0} a_k \log_{10}(\nu)^{k},
\end{equation}
where $a_k$ are coefficients to be fitted \citep[e.g][]{Bowman_edges_2018, SARAS3, Bevins_SARAS3_2022}. In practice, the foreground modelling is not always consistent across different data sets and to further emphasise that the two mock experiments are observing different parts of the sky, we assume they have different complexities and fit a 3-term polynomial to experiment $A$ and a 4-term polynomial to experiment $B$ for the foreground. Generally, when polynomials are being used, the order of the polynomial would be optimised for through an assessment of the Bayesian evidence. We inject Gaussian random noise into our data sets with standard deviations of $35$~mK and $15$~mK for experiments $A$ and $B$ respectively.

\begin{figure*}
    \centering
    \includegraphics{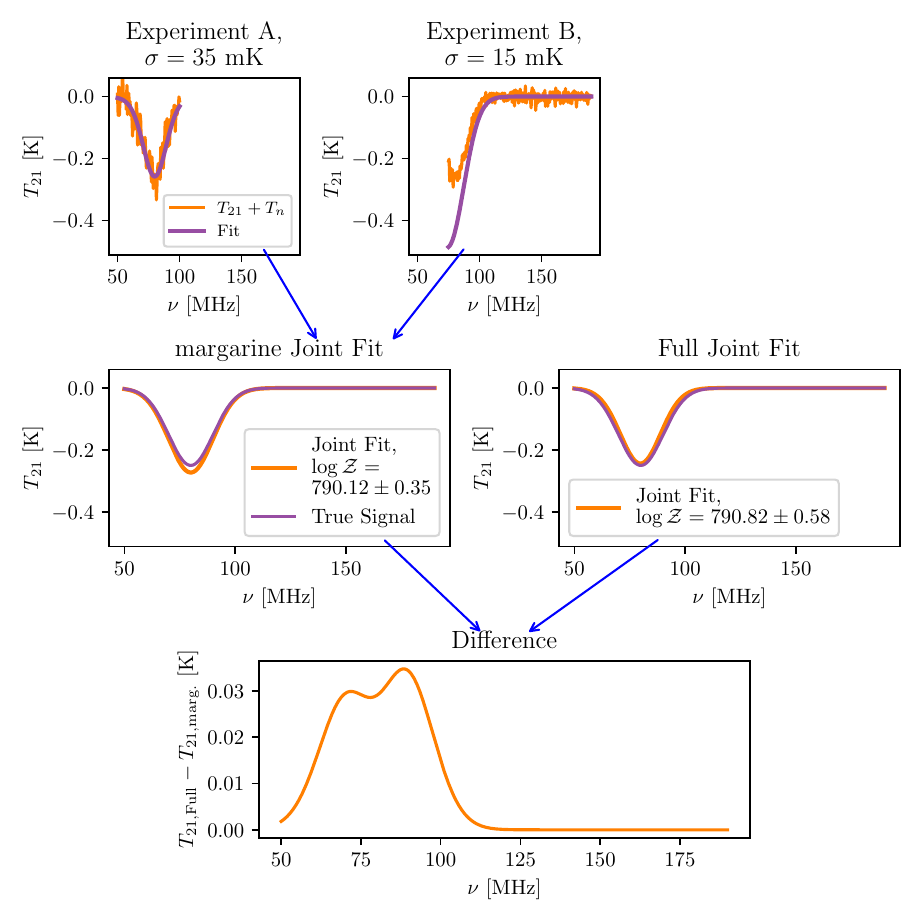}
    \caption{The top panels in the graph shows two simulated data sets comprising noise and a 21-cm signal as orange lines. Note that the data sets also contain foregrounds but we do not show them here since they are several orders of magnitude larger than the 21-cm signal. In the same panels in purple, we show the averages of the functional posteriors derived from Nested Sampling fits to both of these data sets. Neither fit recovers the true signal exactly, however when we analytically combine the full likelihoods for each experiment, including foregrounds, and jointly fit the data we get a much better agreement across the whole frequency range as can be seen in the center right panel. We show the results of this joint analysis as an orange line. For comparison, we show in the center left panel, as an orange line, the average of the functional posterior from a joint fit performed with \textsc{margarine} using the nuisance-free likelihood function. We can see that the joint fits are approximately equivalent, more accurate than the fits to each individual experiment, particularly experiment $B$, and the Bayesian evidences are similar in magnitude when using the full likelihood and the nuisance-free likelihood. We show the difference, which is at most an order of magnitude lower than the expected noise in a global 21-cm experiment, between the two fits in the bottom panel as a function of frequency.}
    \label{fig:joint_likelihood}
\end{figure*}

In our Nested Sampling runs, we use a Gaussian log-likelihood function
\begin{equation}
    \log\mathcal{L} = \sum_i \bigg(-\frac{1}{2}\log(2\pi \sigma^2) -\frac{1}{2}\bigg(\frac{T_\mathrm{D}(\nu^i) - T_\mathrm{fg}(\nu^i) - T_{21}(\nu^i)}{\sigma}\bigg)^2\bigg),
    \label{eq:log_likelihood}
\end{equation}
where $T_\mathrm{D}$ corresponds to the mock data and $\sigma$ corresponds to the instrument noise and is a fitted parameter. In the top row of \cref{fig:joint_likelihood} we show the sum of the noise and signal models that went into the mock datasets and the functional averages of our fitted 21-cm signals
\begin{equation}
    \overline{T}_{21} = \frac{\sum_j^{N_\mathrm{samples}} w^j T_{21}(\theta_{21}^{j})}{\sum_j^{N_\mathrm{samples}} w^j}
\end{equation}
where $w_j$ is the weight corresponding to sample $j$ output by \textsc{polychord}.

We also show in the center right panel of \cref{fig:joint_likelihood} the resultant $\overline{T}_{21}$ found fitting both data sets simultaneously with
\begin{equation}
    \begin{split}
    \log \mathcal{L}_\mathrm{A,B} (\theta, \alpha_{A}, \alpha_B) = & \log \mathcal{L}_\mathrm{A} (\theta, \alpha_A)~+ \\ & \log \mathcal{L}_\mathrm{B} (\theta, \alpha_B),
    \end{split}
    \label{eq:full-likelihood}
\end{equation}
where $\alpha_A$ and $\alpha_B$ are the foreground parameters used to model each data set. Through the combination of the two data sets we can see by visual inspection that we get a much better fit to the data than is achieved for each experiment individually, as is to be expected. 

However, in order to do this we have had to sample over the nuisance parameters modelling the foregrounds, $\alpha_A$ and $\alpha_B$. With \textsc{margarine} we can train density estimators on the posteriors from the individual fits for the core science parameters $\theta = \{T_\mathrm{min}, \nu_c, \Delta \nu\}$ for both experiments and given we know the Bayesian evidence from the previous runs can generate nuisance-free likelihood functions for each experiment as in \cref{sec:nuisance-free-likelihood}. We can then sample
\begin{equation}
    \log \mathcal{L}_\mathrm{A,B} (\theta) = \log \mathcal{L}_\mathrm{A} (\theta)~+ \log \mathcal{L}_\mathrm{B} (\theta),
\end{equation}
over just the parameters of our 21-cm model without having to consider modelling the foregrounds. Since each experiment is independent of the others nuisance parameters, this approach will produce the same results as sampling \cref{eq:full-likelihood}. In practice, we could emulate $\mathcal{P}_\mathrm{A}(\theta)$ and the equivalent for experiment $B$ and perform the joint analysis sampling over
\begin{equation}
    \log \mathcal{P}_\mathrm{A,B} (\theta) = \log \mathcal{P}_\mathrm{A} (\theta)~+ \log \mathcal{P}_\mathrm{B} (\theta),
\end{equation}
and we would arrive at the same posterior distribution found when sampling over $\mathcal{L}_\mathrm{A, B}(\theta)$, however we would not recover the same Bayesian evidence as found when sampling over $\mathcal{L}_\mathrm{A, B}(\theta, \alpha)$.

We show the resultant $\overline{T}_{21}$ from the joint analysis with \textsc{margarine} in the center left panel of \cref{fig:joint_likelihood} and we can see that the fit is consistent with the more complex analysis. In particular, we note the approximate consistency between the Bayesian evidences for each fit. The error on the evidence for the \textsc{margarine} fit is likely underestimated due to uncertainty in the density estimation, however this is currently hard to quantify and left for future work.

In this illustrative example there are only a few nuisance parameters and the likelihood is analytic and quick to evaluate. However, generally speaking there are many more nuisance parameters that need to be modelled and the complexity of the models can lead to significant increases in the evaluation time per call to the likelihood \citep[e.g.][]{Bevins_SARAS2_2022}. In these circumstances, performing marginal joint analysis with \textsc{margarine} can be significantly more efficient than sampling all over the core science and nuisance parameters. This has previously been demonstrated with data from Planck and the Dark Energy Survey \citep{margarine_maxent} and more recently with data from HERA and SARAS3 \citep{Bevins_SARAS3_HERA_2023}.

\subsection{Prior Emulation}

Since the 21-cm signal is dependent on many different astrophysical process, semi-numerical simulations can take several hours to produce a single model. This is impractical if we want to use Bayesian inference techniques like MCMC and Nested Sampling algorithms to fit real data. Signal emulators are often used as a fast alternative that can accurately model the relationship between the astrophysics and the signal structure and produce signal realizations in 10s of milliseconds \citep{Cohen_21cmGEM_2020, Bye_21cmVAE_2022, globalemu}. An even cheaper alternative to emulators is to approximate the signal with a Gaussian absorption feature, as alluded to in the previous section and demonstrated in \cite{Bernardi_2016} and \cite{Monsalve_2017}. However, it is not always clear how to set the priors on the parameters $T_\mathrm{min}$, $\nu_c$ and $\Delta \nu$ in \cref{eq:Gaussian_signal}. Whilst there is some uncertainty in the theoretical modelling of the 21-cm signal, that can be overcome with this phenomenological model, we can motivate our priors on $T_\mathrm{min}$, $\nu_c$ and $\Delta \nu$ with semi-numerical simulations using \textsc{margarine}.

The idea here is to take a large set of physically motivated signals generated using an emulator from a uniformly sampled prior on the astrophysical parameters describing the underlying semi-numerical simulation. For each physical signal we then calculate the depth corresponding to $T_\mathrm{min}$, the central frequency $\nu_c$ and approximate the width $\Delta \nu$. Through this process we arrive at a physically motivated set of samples in the phenomenological parameters which we can train a MAF on using \textsc{margarine}, creating a physically motivated prior on our phenomenological parameters. The resultant model and prior combination is less constrained by the assumptions that go into our semi-numerical models, but still satisfies our broad expectations about the 21-cm signal.

We use the global 21-cm signal emulator \textsc{globalemu} \citep{globalemu} to model a set of 25,000 signals based on a parameterisation of the signal that includes $V_c$ the minimum virial circular velocity that is proportional to the cube root of the minimum halo mass for star formation, $f_*$ the star formation efficiency, $f_X$ the X-ray production efficiency which is proportional to the X-ray luminosity, $\tau$ the CMB optical depth and $f_r$ the radio production efficiency proportional to the radio luminosity. The parameterisation is discussed in more detail in \cite{Reis2020} and references therein. %
Whilst 25,000 signals may seem like a large number of models, the equivalent number of calls made to the emulator in a Nested Sampling or MCMC run would be several orders of magnitude larger.

In scenarios with late and inefficient star formation, corresponding to high $V_c$ and low $f_*$, and strong X-ray heating corresponding to high $f_X$ in the early universe, we find that the signal cannot be well approximated by a Gaussian profile because it features a weak absorption feature and strong emission. We therefore filter out these scenarios before training our MAF and of the 25,000 models we initially started with, 88\% are used to produce our physically motivated prior.

\begin{figure*}
    \centering
    \includegraphics{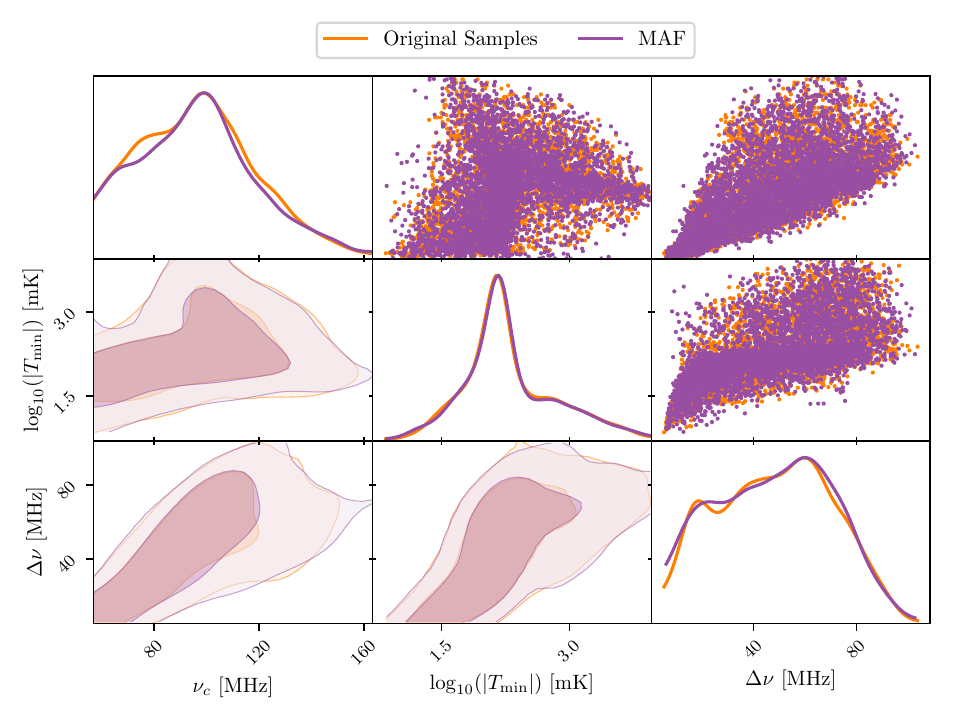}
    \caption{A physically motivated prior on the phenomenological parameters of a Gaussian absorption feature used to model the sky-averaged 21-cm signal. $\nu_c$ is the central frequency of the absorption feature, $T_\mathrm{min}$ is the corresponding temperature, and $\Delta \nu$ is the approximate width of the 21-cm signal. The prior is derived by approximating the phenomenological parameters for a set of physically motivated signals generated with a neural network emulator and learning the corresponding distribution, in orange, with a MAF. Samples from the MAF are shown in purple. The resultant prior and model combination is semi-independent of the assumptions that go into the semi-numerical simulations that the emulator is trained on and quicker to evaluate than the emulator.}
    \label{fig:physical_prior}
\end{figure*}

In \cref{fig:physical_prior} we show the distribution on $\log_{10}(|T_\mathrm{min}|)$, $\nu_c$ and $\Delta \nu$ generated from the physical models and samples from the corresponding MAF which can be used as a prior on the Gaussian model.

In 21-cm cosmology, instrumental effects can introduce non-smooth structure to the data. These effects are often sinusoidal \citep[e.g.][]{Hills2018, Sims2020, Singh_edges_2019, bevins_maxsmooth_2021, Bevins_SARAS2_2022} and if not corrected for they can affect the recovery of the 21-cm signal. For example, part of a damped sinusoidal structure in the data with a periodicity of 5~MHz can be modelled with the approximate Gaussian signal profile discussed in this paper if the prior on the width is wide. However, from \cref{fig:physical_prior} we see that a signal profile of width 5~MHz is not likely to represent a physical scenario. The goal of the physically motivated prior is to therefore provide some information about the anticipated signal that can help to separate it from and prevent fitting systematic structures in the data.

\subsection{Marginal Bayesian Statistics}

Next, we turn our attention to the calculation of marginal Bayesian statistics with \textsc{margarine}.

\subsubsection{Analytic Examples}

First, we investigate a simple two-dimensional multivariate normal distribution with means of zero and standard deviations of one with an analytic KL divergence and BMD. This should be trivial for both the MAF and the KDE to learn, as it resembles the standard normal distribution used for the base of the MAF and the kernel used for the KDE. The analytic KL divergence for this example is given in \cite{Handley_dimensionality_2019} as 
\begin{equation}
    \mathcal{D} = \log \frac{V}{\sqrt{|2\pi \exp(1) \Sigma|}},
    \label{eq:gauss_kl}
\end{equation}
where $V$ is the prior volume and $\Sigma$ is the covariance matrix of the multivariate normal distribution. For a multivariate Gaussian the Bayesian Model Dimensionality is equal to the number of dimensions.

We draw samples from this multivariate normal and train both a MAF and a KDE on the distribution. The calculated statistics are shown in \cref{tab:analytic_examples} assuming a uniform prior that encapsulates the range of the samples. Both the KDE and the MAF produce estimates of the KL divergence in close agreement with the analytic solution. The MAF performs much better when estimating the BMD and the true analytic BMD is within the one sigma error.

The errors in $\mathcal{D}$ and $d$ are estimated by propagating both samples generated with the density estimator and the original samples through the density estimator to evaluate the log-probabilities and comparing the resultant statistics. If the density estimator is a perfect representation of the original distribution, then we would expect samples drawn from it to be from the same distribution as the original samples. This would lead to an approximate equivalence between the two log-probability distributions as functions of the associated Nested Sampling weights~\citep{Harrison2015} and so any deviation we see gives us an indication of the error in our marginal statistics.

In addition to the multivariate Gaussian example, we look at another two-dimensional distribution with an analytic solution where the `posterior' for one parameter follows a triangle distribution and the other is unconstrained following a uniform distribution. In this case, the KL divergence is given as $\mathcal{D} = \log 2 -0.5$ and the BMD as $d = 0.5$. We see a good level of agreement between the analytic KL and BMD estimates from \textsc{margarine}, with both the MAF and KDE assuming a uniform prior as before. Again, however, it appears that the MAF performs better when estimating the BMD in comparison to the KDE.

\begin{table}
    \centering
    \begin{tabular}{|c|c|c|}
         \multicolumn{3}{|c|}{Multimodal Gaussian} \\
         \hline
         & $\mathcal{D}$ & $d$ \\
         \hline
         \hline
         Analytic & 3.134 & 2.000 \\
         \hline
         \textsc{margarine}-MAF & $3.180^{+0.009}_{-0.009}$ & $1.987^{+0.044}_{-0.037}$ \\
         \hline
         \textsc{margarine}-KDE & $3.165^{+0.016}_{-0.017}$ & $1.801^{+0.077}_{-0.068}$ \\
         \hline
         
         \multicolumn{3}{|c|}{Triangle-Uniform} \\
         \hline
         & $\mathcal{D}$ & $d$ \\
         \hline
         \hline
         Analytic & 0.193 &  0.500 \\
         \hline
         \textsc{margarine}-MAF & $0.186^{+0.005}_{-0.005}$ & $0.509^{+0.012}_{0.011}$ \\
         \hline
         \textsc{margarine}-KDE &  $0.181^{+0.014}_{-0.014}$& $0.450^{+0.034}_{-0.029}$ \\
         \hline
    \end{tabular}
    \caption{The table shows the KL divergence and BMD estimated by \textsc{margarine} for two trivial two-dimensional distributions, with analytic KL and BMD values for comparison. We see a good level of agreement between the analytic values and \textsc{margarine} although the MAF typically performs better when estimating the BMD.}
    \label{tab:analytic_examples}
\end{table}

\subsubsection{Accuracy vs Samples and Dimensions}

Using the multivariate Gaussian example, we next assess how the accuracy of the KL divergence and BMD estimates from \textsc{margarine} scale with the number of samples in the distribution and the number of dimensions. In the previous example, we used a mean of zero and a standard deviation of one, however in this section we randomly assign a mean between -2 and 2 for each dimension and a standard deviation of between 0.1 and 1. We use a diagonal covariance matrix. %While this problem is trivial, we expect most unimodal distributions to have similar structures, i.e. to be shifted and scaled versions of the standard normal distribution, after applying the Gaussianization trick shown in \cref{fig:gaussianized}.

\begin{figure*}
    \centering
    \includegraphics{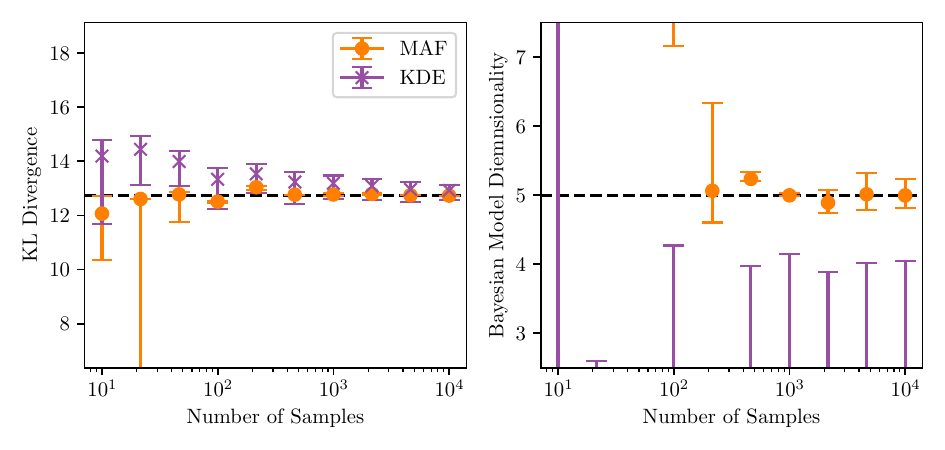}
    \caption{The figure illustrates how the KL divergence and the BMD estimates from \textsc{margarine} converge on the analytic values for a toy example with increasing number of samples in the training distribution. Here the KDE and MAF are being trained on a five dimensional multivariate Gaussian with means randomly chosen from the range -2 to 2 and standard deviations chosen randomly from the range 0.1 to 1. The true KL divergence and BMD shown by the black dashed lines are analytically calculated using \cref{eq:gauss_kl} and equal to the number of dimensions in the distribution respectively. We see that both the KDE and MAF are able to accurately recover the KL divergence for sample sizes $\gtrsim 100$ and the MAF performs well for the BMD with sample sizes $\gtrsim 300$. The KDE significantly underestimates the value of the BMD, however, this is acknowledged by the larger error bars.}
    \label{fig:samples-vs-convergence}
\end{figure*}

\Cref{fig:samples-vs-convergence} shows how the accuracy of the KL estimates, left panel, and BMD, right panel, increase with increasing number of samples for a five dimensional problem. We find that both the KDE and the MAF are able to accurately recover the analytic solution, black dashed lines, for the KL divergence for sample sizes $\gtrsim 100$. We also see that the MAF is more confident and more accurate in its estimate. For sample sizes $\gtrsim 300$, the MAF recovers the BMD well however the KDE consistently underestimates it regardless of the number of samples in the training data. The KDE errors are however larger than for the MAF acknowledging its inability to recover the analytic solution.

\begin{figure*}
    \centering
    \includegraphics{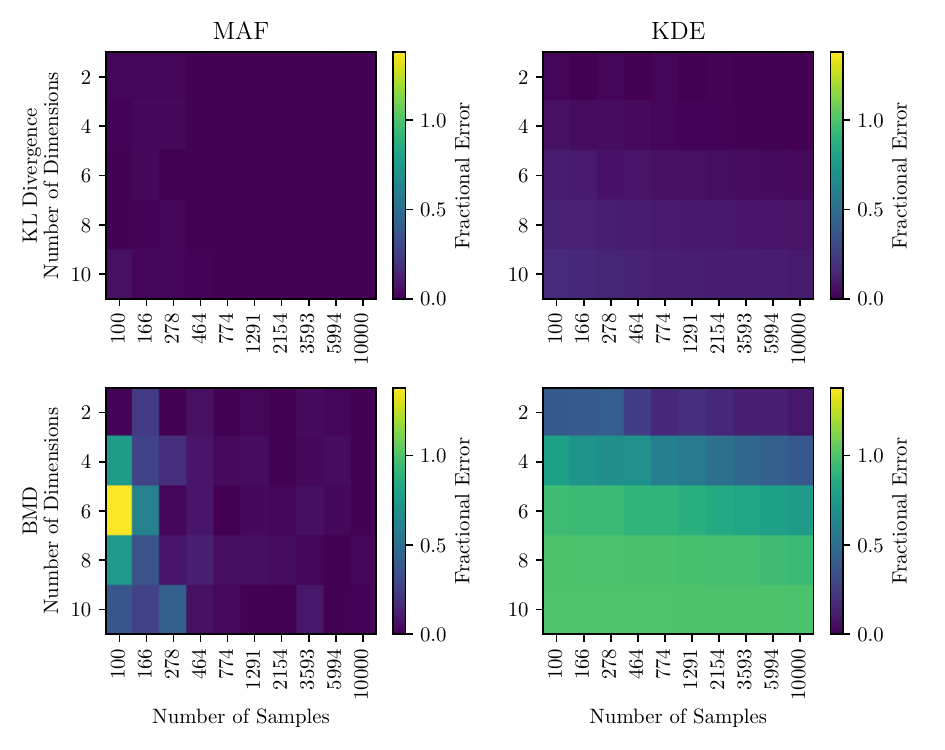}
    \caption{The accuracy of recovered statistics for a multivariate Gaussian distribution with random means and standard deviations per dimension as a function of sample size and dimension. We use \cref{eq:accuracy} to estimate the accuracy and compare the predicted statistics from \textsc{margarine} with the analytic solutions given by \cref{eq:gauss_kl} and the dimensionality of the distributions. The top row shows the KL divergence, the bottom row the BMD, the first column the accuracy for the MAF and the second column the accuracy for the KDE. We see a stronger dependence of the accuracy of the predicted statistics on the number of dimensions and the sample size for the KDE than for the MAF demonstrating that the MAF is much more expressive. We also see a higher level of error for the BMD than for the KL divergence, as previously seen in \cref{fig:samples-vs-convergence}, but note that the fractional error in the MAF BMD estimates is typically much lower than for the KDE. The increased noise across dimension and sample sizes for the MAF is likely due to the random initialisation of the neural network weights.}
    \label{fig:samples-vs-dims}
\end{figure*}

In \Cref{fig:samples-vs-dims}, we explore how the number of samples needs to scale with the number of dimensions for the MAF and KDE to accurately recover the Bayesian statistics. We use the same multivariate Gaussian example with random means and standard deviations but vary the number of dimensions from two to ten. We quantify the level of accuracy using the fractional difference
\begin{equation}
    \epsilon = \frac{|\mathcal{D}_{margarine} - \mathcal{D}_{analytic}|}{\mathcal{D}_{analytic}}.
    \label{eq:accuracy}
\end{equation}
Again it is clear that both the KDE and the MAF are much better at estimating the KL divergence than the BMD likely because the BMD has a stronger dependence on the predicted log-probability and hence on the accuracy of the density estimation. However, the MAF is consistently better at estimating the BMD than the KDE and requires fewer samples than the KDE to accurately recover the BMD.

As expected, when we increase the number of dimensions in the problem, we require more samples to maintain the same level of accuracy. However, this relationship is stronger for the BMD and stronger still for the KDE than the MAF indicating that the MAF is much more expressive 

\subsubsection{Toy Example Posteriors}

\begin{figure*}
    \centering
    \includegraphics[width=\linewidth]{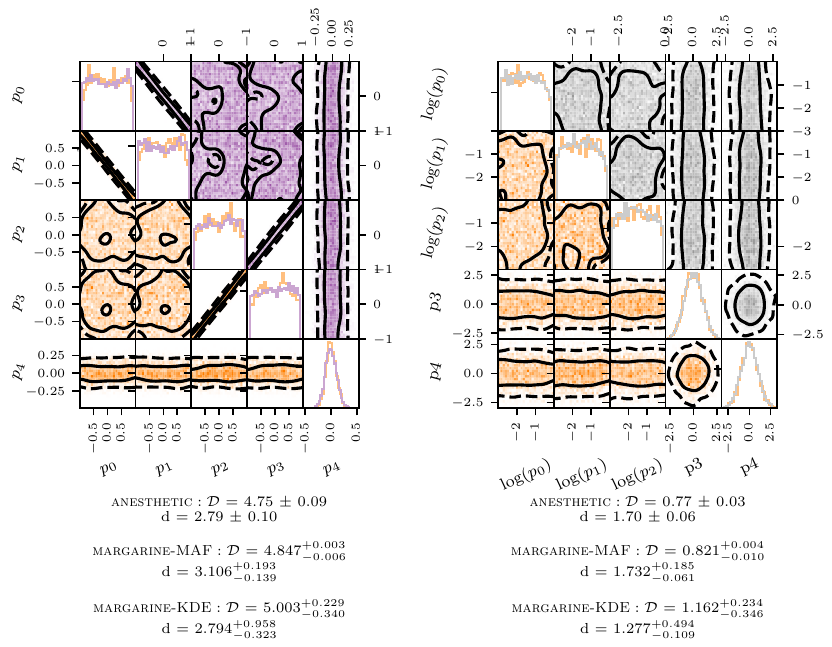}
    \caption{\textbf{Left Panel:} The graph shows a set of correlated Nested Samples from \textsc{polychord} in orange along with a reconstruction from a trained KDE in purple. Listed are the corresponding `true' values for the KL divergence and BMD from \textsc{anesthetic} and the estimated values from \textsc{margarine} using both a KDE and MAF. \textbf{Right Panel:} An equivalent graph for a set of uncorrelated samples with both log-uniform and uniform priors on the parameters, shown in orange. The gray samples are taken from a trained MAF, and we report the `true' Bayesian statistics values from \textsc{anesthetic} along with estimates calculated using a MAF and KDE with \textsc{margarine}.}
    \label{fig:toy_example}
\end{figure*}

Following testing on analytic problems, we generate the examples using \textsc{polychord} and a Gaussian likelihood, both with five parameters, and we show the corresponding distributions in orange in \cref{fig:toy_example}. The first (left-hand panel) has a series of correlations between the parameters and involves sampling the likelihood 
\begin{equation}
    \mathcal{L} = \prod_j \frac{1}{\sqrt{2\pi (0.1)^2}} \exp\bigg(-\frac{1}{2}\frac{(p^{0}_{j} + p^{1}_j)^2 + (p^2_j - p^3_j)^2 + (p^4_j)^2}{0.1^2}\bigg),
\end{equation}
where $p^i_j$ corresponds to parameter $i$ sample $j$ assuming a uniform prior between -1 and 1 for each. The second (right-panel of \cref{fig:toy_example}) has a combination of uncorrelated parameters with both log-uniform and uniform priors on the parameters (right-hand panel) and involves sampling a Gaussian likelihood
\begin{equation}
    \mathcal{L} = \prod_j \frac{1}{\sqrt{2\pi}} \exp\bigg(-\frac{1}{2} \sum_i (p^{i}_j)^2\bigg).
\end{equation}

We are able to use the samples from \textsc{polychord} and the analysis tool \textsc{anesthetic}~\citep{anesthetic} to calculate the KL divergence and BMD for both toy examples and these values are reported, with errors, in \cref{fig:toy_example}. We show samples drawn from a KDE trained on the correlated samples and from a MAF trained on the uncorrelated samples, with values of $\mathcal{D}$ and $d$ listed for both types of density estimator.

For the correlated samples, we see that the estimated KL divergence and BMD from the MAF and from the KDE are in close agreement with the `true' value output from \textsc{anesthetic}. This data set is made of $\approx 7000$ samples and is therefore well within the regime where we expect the MAFs and KDEs to provide an accurate KL divergence estimate and the MAF to produce an accurate BMD estimate (see \cref{fig:samples-vs-dims}.
For the uncorrelated samples, we also see that the MAF KL divergence and BMD estimates are in agreement with the \textsc{anesthetic} values. For the KDE, the KL divergence and BMD are less consistent with the values from the MAF and from \textsc{anesthetic}, however we find we have larger errors when using the KDEs likely because they are less expressive than the MAFs. For this example, we have fewer samples, $\approx 4000$ but we still expect the estimates of the statistics to be accurate, with the exception of the KDE BMD estimate.

Typically, the upper and lower bounded range on the BMD estimates are larger than for the KL divergence estimates because it has a more complex dependence on $\log\mathcal{L}$ and therefore on the accuracy of the density estimation as previously discussed. Future work is needed to investigate whether improvements can be made to the BMD estimates by modifying the MAF loss function, network configurations or KDE bandwidths among other tunable parameters.

\subsubsection{REACH}

REACH is an experiment aiming to detect the sky-averaged 21-cm signal in the frequency range $50 - 170$ MHz \citep{Acedo_reach_mission_2022}.

The REACH data analysis pipeline uses Nested Sampling, implemented with \textsc{polychord}, to model the different components of the data. The pipeline has been extensively tested on mock observations~\citep{Anstey_REACH_2021,Anstey_antenna_2022,Cumner_antenna_2021, Scheutwinkel2022a, Scheutwinkel2022b}. We generate mock observational data using a realistic foreground model derived from an all-sky map and inject a Gaussian signal profile with an amplitude of $|T_\mathrm{min}| = 155$~mK, a central frequency of $\nu_c = 85$~MHz and a standard deviation of $\Delta \nu = 15$~MHz.

\begin{figure*}
    \centering
    \includegraphics[width=\linewidth]{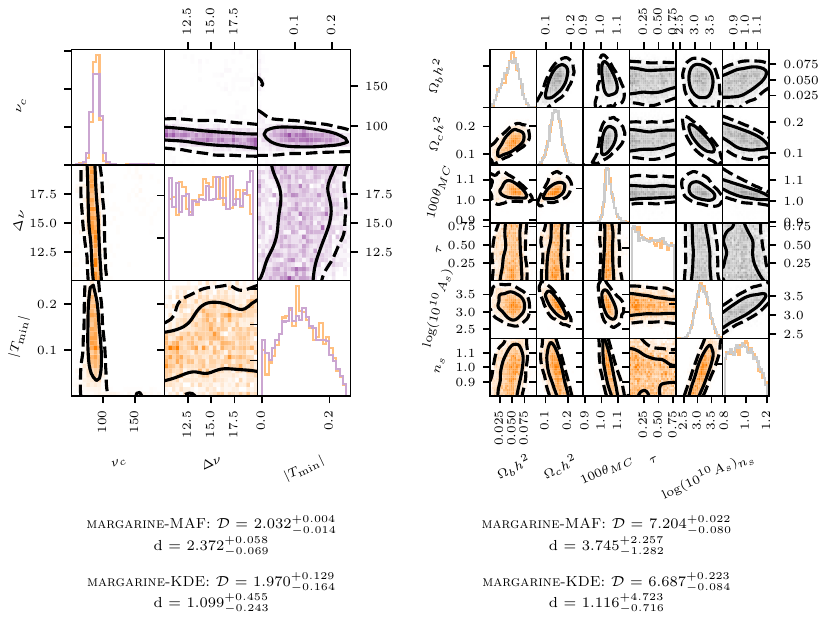}
    \caption{\textbf{Left Panel:} The signal subspace, in orange, for a fit to simulated observations of the 21-cm signal with the REACH experiment along with samples, in purple, from KDE trained on the three signal parameters effectively marginalising over the other thirteen parameters in the fit. We report the marginal KL divergence and BMD found for this set of three parameters using both a MAF and KDE. \textbf{Right Panel:} The cosmological subspace for the Year 1 DES samples shown in orange, along with a set of samples taken from a trained MAF in gray. Again, we report the corresponding marginal Bayesian statistics calculated with \textsc{margarine}.}
    \label{fig:cosmo_examples}
\end{figure*}

The mock data correspond to a single snapshot observation taken from the Karoo radio observatory with a dipole antenna and modelled with a foreground model that takes advantage of the spectral structure of the sky, a correction for the non-uniform response of the antenna to the sky, Gaussian noise and a Gaussian signal profile corresponding to a 16 dimensional parameter space. In the left-hand panel of \cref{fig:cosmo_examples} we show the posterior distribution for the signal parameters from our fit in orange, having marginalised over the other 13 parameters, and the corresponding KDE reconstruction from \textsc{margarine} is shown in purple. We see a reasonable consistency between the marginal KL divergence calculated for these samples when using both the KDE and MAF. 

However, we note that there are some differences in the BMD estimate, with the MAF giving a larger value for $d$. From a visual inspection of the samples, we would expect the BMD to be around 1-2 due to the tight Gaussian constraint on $\nu_c$ and weaker but still non-trivial constraint on $|T_\mathrm{min}|$. Both estimates are, therefore, somewhat consistent with expectations. For this example, we have $\approx 2500$ samples and the levels of accuracy for the MAF and KDE that we see are again consistent with predictions from \cref{fig:samples-vs-dims}.

\subsubsection{Dark Energy Survey}

The Dark Energy Survey~(DES) is designed to help us understand why the Universe's rate of expansion is accelerating. The goal of the collaboration is to map millions of galaxies, thousands of supernova and large scale cosmic structures in order to help understand the dark energy which makes up 70\% of the universe. The survey has targeted the measurements of the dark matter and dark energy densities as well as the dark energy equation of state~\citep{DES_2005, DES_Year1_2018, DES_year3, DES_year3_2021} by investigating galaxy clustering, gravitational lensing and supernova distances.

The Year 1 DES analysis\footnote{We note that the Year 3 DES results have been published~\citep{DES_year3, DES_year3_2021} but at the time of writing that the samples have not been made publicly available.} \citep{will_handley_2020_4116393} aimed to constrain the baryon density parameter, $\Omega_b$, the dark matter density parameter, $\Omega_c$, the approximate ratio of the sound horizon to the angular diameter distance, $\theta_{MC}$, the optical depth of the CMB to reionization, $\tau$, and the amplitude and tilt of the power spectrum, $A_s$, and $n_s$. The samples from the cosmological subspace are shown in orange in the right-hand panel of \cref{fig:cosmo_examples}, where we have marginalised over a set of 20 nuisance parameters, along with a MAF reconstruction of the subspace in gray.

Unlike the toy examples and REACH samples, the priors on the DES analysis are non-uniform, astrophysically informed and cannot easily be transformed into a space in which the priors are uniform. As a result, we have to use the density estimators built into \textsc{margarine} to evaluate both the log-probability of the posterior and the prior if we want to calculate marginal statistics.

While this is possible, it is expected to lead to an increased uncertainty in the marginal statistics, which can be seen in \cref{fig:cosmo_examples}. In addition, the problem is further complicated by the size of the cosmological parameter space, since we expect larger parameter spaces to be harder to replicate well with \textsc{margarine}, and consequently we expect larger errors on the marginal statistics.

Again, the density estimators give us different estimates for the BMD, however the errors are large. Here we have $\approx 7000$ samples and six dimensions, and from \cref{fig:samples-vs-dims} we would expect the BMD estimate to be better from the MAF. From a visual examination of the distribution, we would expect the BMD to be between $\approx3$ and $\approx4$ given the Gaussian nature of the 1D posteriors on $\Omega_b h^2$, $\Omega_c h^2$, $\theta_{MC}$ and $\log(10^{10}A_s)$. There is a reasonable agreement between the KL divergence calculated for the cosmological constraints from DES with both the MAF and the KDE.

\section{Conclusions}
\label{sec:conclusions}

In this paper we have demonstrated a number of applications of two different types of density estimator, Masked Autoregressive Flows and Kernel Density Estimators, to the calculation of marginal Bayesian statistics, the efficient modelling of multiple data sets and the derivation and emulation of physical priors. \textsc{margarine} is a multipurpose tool that can be used to enhance our Bayesian analysis workflows.

The evaluation of marginal KL divergences and Bayesian Model Dimensionalities allows for improved comparison of the constraining power of different experiments targeting the same astrophysical or cosmological signals with different systematics or nuisance parameters. In turn, this can lead to improvements in experimental design with techniques that provide more information about the signals of interest being pursued in the future. The calculation of marginal Bayesian statistics has been illustrated in \cite{Scheutwinkel2022b}, \cite{Anstey2023} and \cite{Bevins_SARAS3_2022}.

We find that the MAFs are much more expressive than the KDEs and perform better when estimating the KL divergences and BMDs. KDEs, however, perform well when estimating the KL divergence in low dimensions with large sample sizes as illustrated in \cref{fig:samples-vs-dims} and are quicker to train than the MAFs. In practice, we would advise the use of MAFs over KDEs because of their higher degree of accuracy and better scaling with dimensions and sample sizes.

The nuisance-free likelihood function allows for more efficient combination of constraints from different data sets, preventing the need to sample instrument specific foreground and systematic effects in joint analysis. This was demonstrated with data from the Planck and Dark Energy Survey experiments in \cite{margarine_maxent} and more recently with data from the 21-cm power spectrum experiment HERA and sky-averaged 21-cm experiment SARAS3 \citep{Bevins_SARAS3_HERA_2023}.

Finally, \textsc{margarine} can be used to generate non-trivial priors \citep[e.g.][]{Alsing_bijectors_2021} either from experimental results or simulations, as illustrated in this paper. In principle, this allows us to inform the analysis of data from upcoming probes like REACH \citep{Acedo_reach_mission_2022} or the Simons Observatory \citep{Simons_Obs_2019, Simons_obs_2019b} with results from existing experiments like SARAS3 \citep{Bevins_SARAS3_2022} and Planck \citep{Planck_cosmo_2020}.

We anticipate further development of \textsc{margarine} and additional unexpected applications that evolve as the code is used \citep[e.g.][]{bevins_pnf_2023}.

\section{Acknowledgements}

HTJB acknowledges the support of the Science and Technology Facilities Council (STFC) through grant number ST/T505997/1 and support from the Kavli Institute for Cosmology, Cambridge and the Kavli Foundation. WJH and AF were supported by Royal Society University Research Fellowships. PHS acknowledges support from a Trottier Space Institute Fellowship and the Canada 150 Research Chairs Program. EdLA was supported by the STFC through the Ernest Rutherford Fellowship. JA was supported by the research project grant “Fundamental Physics from Cosmological Surveys” funded by the Swedish Research Council (VR) under Dnr 2017-04212.

\section{Data Availability}

The DES samples are available at \url{https://zenodo.org/record/4116393#.Y9zfPC8RpKM}. All other data is available upon reasonable request to HTJB.

\bibliographystyle{mnras}
\bibliography{journals, refs}
\bsp	% typesetting comment
\label{lastpage}
\end{document}